\documentclass{aa}
\usepackage{graphicx}
\begin{document}
\title{Absorbers and Globular Cluster Formation in\\
  Powerful High Redshift Radio Galaxies}
\author{M.Krause}
\institute{Landessternwarte K\"onigstuhl, D-69117 Heidelberg, 
  Germany}
\offprints{M.Krause, email: 
{M.Krause@lsw.uni-heidelberg.de}}
\date{Received December 14, 2001 / Accepted <date>}
\begin{abstract}
A radiative hydrodynamic simulation for a typical, powerful high redshift 
radio galaxy is presented.
The jet is injected at one third the speed of light into a 10000 
times denser, homogeneous medium. 
In the beginning of the simulation, the bow shock consists of a 
spherical shell that is similar
to a spherical blast wave. 
This shell cools radiatively down to 
$\approx 10^4$ K, providing after $6 \times 10^6$ yrs
a neutral column of $3.8 \times 10^{21}\,\mathrm{cm}^{-2}$ 
around the whole system. The shell starts to fragment and forms condensations.
This absorbing screen will cover a smaller and smaller
fraction of the radio source, and therefore the emission line region,
and eventually form stars in typically $10^4$
globular clusters of $10^6 \mathrm{M}_\odot$.
Approximately $10^9 \mathrm{M}_\odot$ are entrained into the radio cocoon. 
This gas, cooling and illuminated from the radio source, 
could be the emission line gas observed in high redshifted radio 
galaxies and radio loud quasars. 
The neutral column behind the bow shock can account for the 
absorption found in almost all of the small sources. 
The globular cluster excess 
of $\approx 10^4$ systems found
in present day brightest cluster galaxies (BCGs),
which are believed to be the vestiges of these objects,
is consistent with the presented scenario. \end{abstract}

\authorrunning{M. Krause}
\titlerunning{Absorbers and Globular Clusters in High Z Radio Galaxies}
\maketitle

\section{Introduction}
Collimated radio-luminous outflows are by now detected from the 
cosmological neighborhood up to redshifts in excess of five.
According to the unified scheme these jet-sources come in two classes: 
Radio-loud quasars, where the current redshift record is $z=5.8$
(\cite{F01}), and radio galaxies. Among the latter ones the highest 
known redshift is $z=5.2$ (\cite{vB99}).
Infrared observations indicate that radio galaxies contain a luminous 
quasar hidden by a dusty torus, at least at redshift above $z\approx0.8$
(\cite{M01}). The maximum power of radio galaxies increases with redshift
(\cite{dB00}), the suspicion being that the most powerful active galaxies
at $z>2$ develop into BCGs 
(\cite{C00}, and references
therein). This idea is supported e.g.
by the fact that the co-moving space density of clusters at low redshift
is comparable to that of powerful AGN at $z=2.5$ (\cite{W94}). 

There is good reason to believe that the jets of higher redshift 
sources bore into a medium of higher density. If one just considers 
the cosmological expansion, the number density of the intergalactic
material (IGM) scales like
$n_\mathrm{IGM} \propto (1+z)^{3}$, on large scales.
This means e.g. 
that if the well known radio galaxy \object{Cygnus A}, which has 
a surrounding IGM number density of a few times $10^{-2} \; \mathrm{cm}^{-3}$ 
(\cite{C96}),
was located at redshift $z=3$, its jet had to bore into a medium with 
$n \approx 1 \; \mathrm{cm}^{-3}$. However, the environments of the 
most powerful radio galaxies,
which are believed to highlight the highest density peaks in the early
universe, are not a priori expected to simply follow the Hubble flow.
For example, it has to be taken into account that the associated galaxies 
appear as bright ellipticals only at redshift $z<1.5$, whereas above 
this redshift one finds clumpy regions, distributed over upto 100 kpc
(\cite{P01}), and a considerable fraction of the gas is transformed into 
stars around that redshift. Direct estimates from emission 
line observations
indicate IGM densities in the range of $0.1 \; \mathrm{cm}^{-3}$ 
(\cite{vO97}) upto
$10 \; \mathrm{cm}^{-3}$, 
if one ascribes the dominant energy source for ionization
to the bow shock of the jet (\cite{Bic00}). If the jets themselves consist in
electron proton plasma and are subrelativistic on large scales -- e.g. in 
\object{Cygnus A} a kpc scale 
jet velocity ($v_\mathrm{J}$) 
of 0.4 $c$  is favoured (\cite{C96}, $c$ being the velocity of light) -- 
the number density in the jet ($n_\mathrm{J}$) is given by 
\begin{equation}
n_\mathrm{J}=\frac{\dot{M}_\mathrm{J}}{\pi R^2_\mathrm{J} v_\mathrm{J} m_\mathrm{H}}=1.3 \times 10^{-4} \;{\rm cm}^{-3} \frac{\dot{M}_\mathrm{J} c}{3 v_\mathrm{J}} \left(\frac{\rm kpc}{R_\mathrm{J}}\right)^2.
\end{equation}
Here $\dot{M}_\mathrm{J}$ is the jet's mass flux rate 
(in solar masses per year), $R_\mathrm{J}$ is the jet radius (about 0.6 kpc
in Cygnus A, \cite{C96}), and $m_\mathrm{H}$ the mass of hydrogen. 
Adopting an IGM number density of $1 \,\mathrm{cm}^{-3}$ for high redshift 
radio sources gives a density contrast
$\eta = n_\mathrm{J}/n_{\mathrm{IGM}} \approx 10^{-4}$. Due to the lower 
environment density (\cite{D95}), 
$\eta$ is thought to be higher in low redshift sources.
Because the expansion velocity of the radio structure scales proportional to 
$\sqrt{\eta}$, the average extension of radio galaxies with $z<1.5$ of 100 kpc
versus 10 kpc at $z>2$ (\cite{C00}) could be nicely explained by such a
cosmological IGM density increase, if the average active time stays constant. 
Another indication for high density 
environments at high redshift is the observed bending of the 
kpc scale radio jets
(e.g. \cite{P97}). This underlines the clumpy nature of 
the IGM around high redshift radio galaxies (HZRG).

The optical emission of radio galaxies 
is often aligned with the jet axis. While below $z=1.5$
this emission is highly polarized, indicating mainly scattered light from a 
hidden AGN, above $z=2$ less then 2\% polarization is found and stellar 
absorption lines are detected (\cite{C00}). This could indicate 
that --  at high redshift -- jets induce star formation. 
The aligned light could also reflect 
the large scale matter distribution (\cite{W94}).

Radio galaxies are surrounded by a Ly-$\alpha$ halo. The size of this halo
grows with redshift (\cite{C00}). Whilst sources at $z<1.5$ have halos of
$10^8 \;\mathrm{M}_\odot$ and an extension of 10 kpc, $z>2$ sources
have 100 kpc halos of $10^9 \;\mathrm{M}_\odot$. 
The width of the emission lines is typically $(1000 \pm 500)$ km/s 
(\cite{vO97}; \cite{dB00}). HZRG with diameters 
less than 50 kpc show associated narrow absorption lines with a width of
$(\approx 40 \pm 30)$ km/s (\cite{vO97}). The absorption is often saturated 
with column densities of $(10^{18} - 10^{20}) \,\mathrm{cm}^{-2}$, and is 
preferentially blue-shifted by upto 250 km/s. These absorption systems
were thought of to consist of $\approx 10^{12}$ solar system sized clouds 
by van Ojik et al. (1997). Advanced modeling with photoionization codes
lead to the proposition that the absorber was a low density region,
surrounding every HZRG (\cite{Bin00}).
None of these models take into account the hydrodynamic facet of the problem.
This letter is therefore dedicated to an understanding of the above 
mentioned features based on hydrodynamic modeling of HZRGs.

\section{A hydrodynamic model for high redshift radio galaxies} 
One of the basic differences between jets at low and high 
redshift is the environmental
density. We consider a -- typical --  
jet with a density contrast of $\eta=10^{-4}$, a velocity
of $v_\mathrm{J}=10^{10}$ cm/s (internal Mach number 85), 
a jet radius of 1 kpc, 
and a density of the homogeneous external medium of 1 $\mathrm{cm}^{-3}$.
This results in a kinetic jet luminosity of $10^{45.4}$ erg/s, which is 
transfered by some efficiency factor to radio luminosity. The velocity
of advancement of the jet's head -- 
and the bow shock in front of it -- is expected 
to be $v_\mathrm{B}\approx \sqrt{\eta} \;v_\mathrm{J} = 1000$ km/s.
This velocity causes the external gas to get heated to a temperature of
$T=1.38 \times 10^7 \;\mathrm{K} \;(v_\mathrm{B}/1000 \,\mathrm{km/s})^2$, 
according to standard hydrodynamic shock jump conditions, neglecting 
the internal energy of the pre-shock gas. 
The cooling time of this 
post-shock gas due to thermal bremsstrahlung with an emissivity of
$\epsilon_\mathrm{brems}=2.1\times10^{-27}(n/\mathrm{cm}^{-3})^2\sqrt{T/\mathrm{K}}$ 
$\mathrm{erg} \;\mathrm{cm}^{-3} \,\mathrm{s}^{-1}$ will be:
\begin{equation}
t_\mathrm{cool}\simeq \frac{2 n k T}{\epsilon_\mathrm{brems}}\approx 3 \times 10^6  \frac{\sqrt{T/10^7\;\mathrm{K}}}{n/4\;\mathrm{cm}^{-3}} \, \mathrm{yrs},
\end{equation}
which is less than the typical lifetime of a jet. Therefore, cooling 
has a major effect on the evolution of the post-shock gas. This gas will 
cool down to high densities and a temperature of about
$10^4$ K, where the cooling curve drops significantly (\cite{SD93}).
\begin{figure}
\centering
\resizebox{.8\hsize}{!}{\rotatebox{0}{\includegraphics{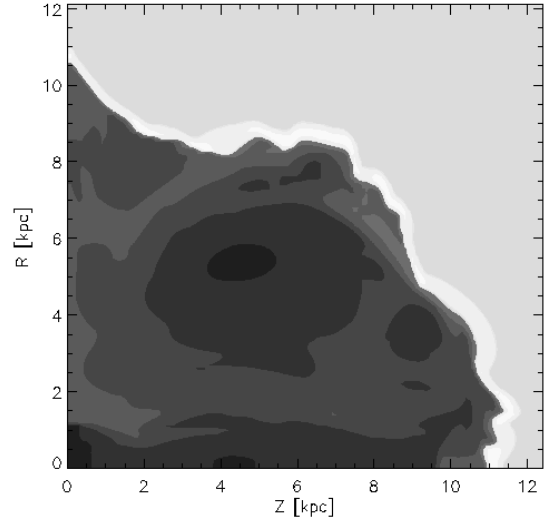}}}
 \caption{Grayscale plot of the simulation described in the text. The levels
   show increasing number density with a factor of ten between the steps 
   (one additional level is drawn at $10^{0.1}\;\mathrm{cm}^{-3}$ in order to 
   highlight the shell). The simulation time is 6.0 mio yrs.}
\label{den310}
\end{figure}
\begin{figure}
\centering
\resizebox{\hsize}{!}{\rotatebox{-90}{ \includegraphics{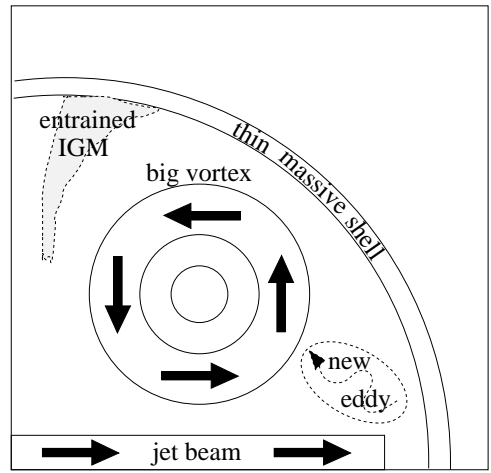}}}
 \caption{Sketch of the basic results of the simulation.}
\label{sketch}
\end{figure}
We have solved the hydrodynamic equations for the above mentioned 
parameters numerically with the code {\em NIRVANA\_C} (\cite{ZY97}),
upgraded due to optically thin cooling (\cite{T00}).
NIRVANA\_C is a second order accurate finite difference scheme.
Cooling comes into play via the equation for the internal energy 
density ($\cal{E}$):
\begin{equation}
\frac{\partial{\cal E}}{\partial t}+\nabla \cdot({\cal E}v)=-p(\nabla \cdot v)
-\Lambda,
\end{equation} 
where $\Lambda$ is the global cooling function. NIRVANA\_C also solves
the ionization and recombination equations, by a time implicit method,
in order to allow for nonequilibrium processes.
For the present simulation, three species were used, namely ionized and 
neutral hydrogen, and electrons. Below $T=10^6$ K, an average cooling curve
for solar metalicity was applied (\cite{SN93}).
The simulation was carried out in axisymmetry (2.5D), and the jet is resolved
with 10 points. With that resolution, NIRVANA\_C marginally resolves shocks
in the jet beam and gets basic flow parameters right. 
Turbulence on small scales as well as details of the beam, especially
of the head region, are unresolved (\cite{KC01}). The simulation was stopped
when the thickness of the shell behind the bow shock shrinked below
the resolution, which was clearly the case after a simulation time of 
$6.0 \times 10^6$ yrs.

The results are quite unusual compared to simulations of higher $\eta$:
Most of the mass accumulates into an almost spherical shell behind the
bow shock. The maximum density there is $47\;\mathrm{cm}^{-3}$, and the gas 
has cooled to an average temperature of $1.7 \times10^4$ K in regions 
with density above $10\;\mathrm{cm}^{-3}$, which still cover almost the whole
shell surface. The mass accumulated in this shell is 
$6.19\times10^{10}\;\mathrm{M}_\odot$, 75~\% of which are neutral. 
This gives an
average neutral column for the shell of 
$N_\mathrm{HI}=3.8\times10^{21}\;\mathrm{cm}^{-2}$, which
has to be compared to the amount of matter displaced by the shell:  
$6.32\times10^{10}\;\mathrm{M}_\odot$. It follows, that 
98 \% of the mass are accumulated in the thin shell. Therefore 2 \% of the
mass or $1.4\times10^{9}\;\mathrm{M}_\odot$ were entrained into the cocoon.

The cocoon itself has a very peculiar structure. It is well known, that 
jet cocoons broaden when lowering $\eta$ (\cite{N83}). But in this simulation,
the cocoon is not elongated along the jet beam, forming 
individual vortices. Instead, all the vortices 
join into one big vortex with a diameter of roughly 8 kpc.
The vortex shedding is sometimes violent: while, at the time shown,
only a small eddy is ejected at the beam head, at other times we observed
upto half the jet beam splitting and forming an eddy.
The big vortex turns around at a maximum velocity of 
$\approx 0.2-0.25\, v_\mathrm{J}$. 

The velocity of the jet head differs considerably from higher $\eta$
simulations. At early times, it is about 10 times higher than the expectation
above. Only after about two million years, the jet head moves, relatively
constant, at maximum expected pace.

Within the cocoon, the sound speed is, on average, 20000 km/s.
Therefore, pressure differences within the cocoon gas are rapidly 
communicated, with the result that the pressure is almost constant. 
The simulation shows that the bow shock is, at early times, perfectly 
spherical. It looks more like a supernova bubble than a jet cocoon.
Indeed, if one compares the velocity of a spherical blast wave 
(e.g. \cite{WP01}, with constant energy ejection):
$v_\mathrm{BW}=1412 
  \left(\frac{L/10^{46}\,\mathrm{erg/s}}{n/\mathrm{cm}^{-3}}\right)^{1/5}
  \left(\frac{t}{\mathrm{10^6\,yrs}}\right)^{-2/5}\mathrm{km/s}$
to the velocity of the jet head, one finds, that the jet head outruns
the blast wave only at times above $1.9 \times 10^6$  yrs, 
for the parameters of our 
simulation. This is what we observe. 
We rewrite the above equation for the case, when the energy injection
comes from a jet:
$v_\mathrm{BW}/v_\mathrm{J}=0.6 \;(R_\mathrm{J}/R_\mathrm{BW})^{2/3} \eta^{1/3}$.
The blast wave velocity depends on $\eta^{1/3}$, whereas the jet head velocity
depends on $\eta^{1/2}$ (light jet limit). Because of that, the spherical 
state of the bow shock can be observed only at the earliest times, for higher 
$\eta$. 

\section{Comparison to observations}  
Because of the large neutral column, this shell is a good candidate 
for the absorber in HZRGs. The width of an absorption 
line caused by this absorber would be approximately the sound speed within it
(\cite{D80}),
which is in our simulation about 55 km/s, on average.
This is also in remarkable agreement with observations (\cite{vO97}).
Typically, the material that is just heated by the shell emits Lyman $\alpha$
at a few percent of the kinetic jet luminosity. This is less
then the observed Ly $\alpha$ luminosity, indicating that the bow shock is not 
the main emitter.
At late times compared to $t_\mathrm{cool}$, it is justified to approximate 
the bow shock with an isothermal shock description. Following Steffen et al.
(1997), we derive for the thickness of the spherical thin shell:
\begin{equation}
d=27.7 \frac{(\bar{m}/m_\mathrm{p})(r/10\,\mathrm{kpc})(T/10^4\,\mathrm{K})}{(v_\mathrm{bw}/100\,\mathrm{km/s})^{2}}\, \mathrm{pc}.
\end{equation}
Here, $\bar{m}/m_\mathrm{p}$ denotes the mean molecular weight per proton mass.
The sound crossing time through this shell is about $5 \times 10^5$ yrs,
which is much longer than the cooling time (about $10^4$ yrs, \cite{SD93}).
Therefore, it is possible, that the shell breaks up into individual subshells
(\cite{D80}). This could well correspond to the multiple associated
absorption systems, often observed in HZRGs.
These absorbers are preferentially blue shifted, a natural feature of our 
model. But almost never, a velocity above 250 km/s is observed. 
Approximating the sideways expansion 
of the shell by the laws for the spherical shell
also for some time after the jet head begins to shape the shell, 
one can calculate $\eta$, given the bubble size ($>12$ kpc) at observation time.
This yields $\eta=2 \times 10^{-5}$, which means that the full range of
currently known observations can be explained if the environmental density 
is $\approx 5 \;\mathrm{cm}^{-3}$. $\eta$ in the above formula could be replaced by
$\eta_\mathrm{0} (R/R_\mathrm{J})^{-p}$ to account for a declining density distribution.
The $\eta_\mathrm{0}$ would be lower, in that case.
It is interesting to note here, that very similar numbers have been found 
from quite different arguments for the environment of the radio galaxy 4C 41.17
(\cite{Bic00}). 

The bubble interior looses a considerable 
amount of internal energy via synchrotron radiation, which is not included
in the simulation. This lowers the requirement for the external density. 
Although the observed hydrogen mass in these systems barly exceeds 
$10^9 \mathrm{M}_\odot$, it is likely that they contain significantly 
more mass. Given that HZRG host the most powerful quasars of the universe,
we should expect black hole masses of the order $10^9 \mathrm{M}_\odot$.
The gas mass should be more than that. If the gas mass would be of the order
$10^{11-12} \mathrm{M}_\odot$, as is suggested here, the gas to black hole mass 
ratio would be similar to the bulge to black hole mass ratio found at low 
redshift. In that case, most of the shell's mass would be -- at any 
evolutionary state, the systems are observed at so far --
in cold fragments just about to form stars.

The shell is not only thermally, but also gravitationally unstable.
The time, when the instability on the shell surface occurs for the first
time is (\cite{WP01}, again for constant energy injection rate and
for the parameters of our model):
$t_\mathrm{i}=2.4 \times 10^7\left(\frac{(c_\mathrm{s}/55\,\mathrm{km/s})^5}{(r/\mathrm{kpc})^2 (\eta/10^{-4}) (3v_\mathrm{J}/c)^3 (n/\mathrm{cm}^{-3})^5}\right)^{1/8}\mathrm{yrs}$.
Gravitational and thermal instability will support each other. 
The Jeans mass in the shell
is $\approx 10^6-10^7 M_\odot$. Hence, the shell will form stars 
in globular clusters of that mass. Assuming an efficiency of $10\%$,
the shell would form, roughly, $10^4$ globular cluster systems.
This is in remarkable agreement to what is found in nearby BCGs, like e.g.
\object{M87} (\cite{H98}): BCGs show an excess of globular cluster systems
of about $10^4-10^5$ compared to non BCG ellipticals with comparable 
luminosity. Sometimes, 
two distinguished globular cluster populations are observed.
Harris et al. (1998) consider various formation scenarios in detail and 
conclude that a kind of galactic wind must have driven out a large fraction
of the galaxies gas, before it was able to form stars.

When the shell fragments, its covering factor decreases further and the absorption 
vanishes. 
This scenario could nicely explain, why no absorbers are observed
for galaxies larger than 50 kpc. Afterwards the shell would become
visible in the optical, due to the newly born stars, which will also increase the ionization of the
remaining parts of the absorbing shell. 
A bubble shaped star forming region like that can be found around the radio 
galaxy \object{1243+036} (\cite{vO96}). 

These associated absorption systems also seem to be 
observed in high redshift quasars
(\cite{B97}; \cite{B01}). There, they have velocities of a few thousand
km/s, consistent with the model, if one assumes that the line of sight 
in quasars is not far from the jet axis. In addition, this part of the shell
could be accelerated by photoabsorption of photons from the quasar 
(\cite{F81}).

In the simulation, we find that $\approx 10^9 M_\odot$ of shocked
IGM are entrained into the cocoon. This number should not be taken too 
literally, because small scale turbulence could change that
and high resolution studies are required to get this number accurate.
Nevertheless, it is slightly more than the typical 
observed Lyman $\alpha$ emitting gas mass. Higher resolution would also
be essential in order 
to determine the mixing properties and density of this gas. 
The big vortex, we observe in the cocoon is an ideal accelerator
for the entrained gas. In the simulation, the gas is swept along,
and accumulates mainly on the left-hand side and along and next to the 
jet beam. There it is spun up to a velocity of $\approx 1000$ km/s.
We expect, that at higher resolution, small scale Kelvin-Helmholtz 
instabilities will entrain mass at the boundary of the vortex.
This is also observed in other simulations (\cite{KC01}).
Given the high velocity of this propeller, it seems quite likely 
that emission line gas could be accelerated to the observed velocity 
of $\approx$ 1000 km/s. All that should be 
studied in more detail with simulations of higher resolution.

\begin{acknowledgements}
This work was supported by the 
Deutsche Forschungsgemeinschaft
(Sonderforschungsbereich 437).
\end{acknowledgements}

{}
\end{document}